# Fano Resonances in High-$T_c$ Superconducting Metamaterial

V.A. Fedotov[1,*], J.-H. Shi[1], A. Tsiatmas[1], P. de Groot[2], Y. Chen[3] and N.I. Zheludev[1]

[1]*Optoelectronics Research Centre, University of Southampton, SO17 1BJ, UK*
[2]*School of Physics and Astronomy, University of Southampton, SO17 1BJ, UK*
[3]*Rutherford Appleton Laboratory, Chilton, Didcot, Oxon, OX11 0QX, UK*
[*]*Email: vaf@orc.soton.ac.uk, web: http://www.metamaterials.org.uk*

**Abstract:** We demonstrate a millimeter-wave range metamaterial fabricated from cuprate superconductor. Two complementary metamaterial structures have been studied, which exhibit Fano resonances emerging from the collective excitation of interacting magnetic and electric dipole modes.

Our interest in superconducting metamaterials is driven by the desire to develop low loss media supporting high quality resonances. Such resonances may be achieved in metamaterials with broken structural symmetry supporting Fano resonances [1] where the quality factor can be controlled by design and is only limited by the Joule losses. Cuprate superconductors show lower surface conductivity than copper at frequencies below 200 GHz even at liquid nitrogen temperatures (see Fig. 1a) and is a prime choice for developing such metamaterials. Here we report the first experimental data on observation of Fano resonances in superconducting metamaterials and demonstrate that their quality factor may be controlled by temperature. All our measurements were performed using a free-space setup, which is based on mm-wave test system equipped with horn antennas (see Fig. 1b) and liquid helium cryostat.

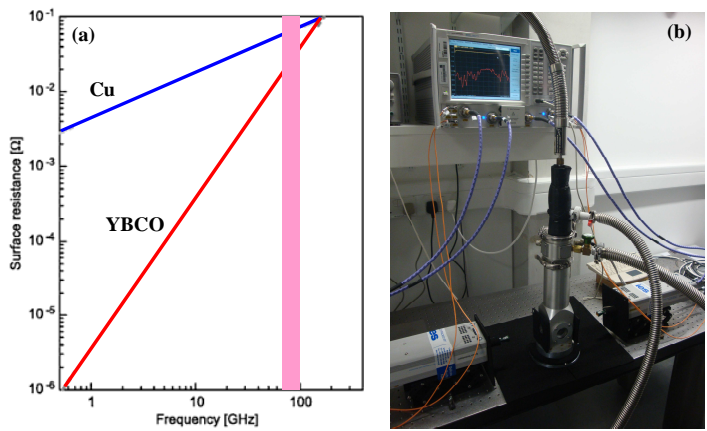

Fig. 1. (a) Surface resistance of copper and YBCO as a function of frequency at 77K. Pink area corresponds to the spectral region covered by our mm-wave experimental setup (shown on panel (b)).

Our metamaterials were fabricated by etching arrays of both positive and negative forms of asymmetrically-split rings in 330 um thick film of high-temperature superconductor YBCO deposited on a low-loss sapphire substrate, as show in Figs. 2a and 2b. Electromagnetic properties of the superconducting structures were studied at temperatures above and below the critical temperature $T_c$ = 87.4 K in 75 – 110 GHz range of frequencies. Our measurements clearly showed the appearance of the Fano resonances upon superconducting phase transition. The results of the measurements are presented on Figs. 2c and 2d, where we plot changes in the transmission spectra of the cuprate metamaterials with decreasing temperature (down to 77 K) relative to their room temperature state. Fano resonances in metamaterials with broken structural symmetry appear as a result of excitation of the so-called trapped mode (electromagnetic mode that is weakly coupled to free-space [1]), and can be seen to fully develop in our cuprate structures at around 87 GHz with temperature dropping below $T_c$. In the case of positive superconducting metamaterial the resonance emerges as a peak in the transmission change, corresponding to increased transmission, while the complementary (i.e. negative) version of the structure shows a pronounced dip at 87 GHz corresponding to resonantly suppressed transmission (which fully agrees with the Babinet principle).

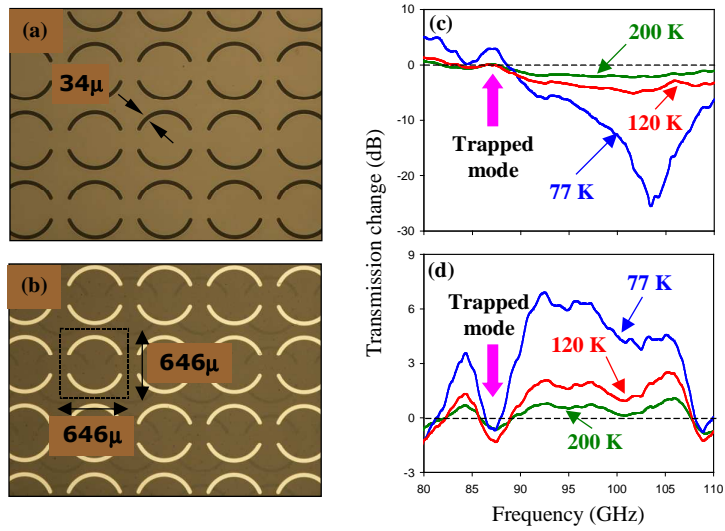

Fig. 2. Superconducting metamaterial. Panels (a) and (b) show positive and negative types of the metamaterial array composed of asymmetrically-split ring metamolecules (as indicated by the dashed box) etched in a form of correspondingly lines and slits in a thin film of high-Tc superconductor YBCO. Panels (c) and (d) show temperature-driven changes in the spectrum of correspondingly positive and negative types of the metamaterial metamaterials with respect to 300K.